\documentclass[11pt,a4paper]{article}

\usepackage[T1]{fontenc}

\usepackage{graphicx}

\usepackage{amsthm}

\usepackage[T1]{fontenc}
\usepackage{color}
\usepackage{amsmath}
\usepackage{amssymb}
\usepackage{subcaption}
\usepackage{tikz}
\usepackage[ruled, linesnumbered, vlined]{algorithm2e}

\newcommand{\tbf}[1]{\textbf{#1}}
\newcommand{\mcal}[1]{\mathcal{#1}}

\newcommand{\mrm}[1]{\mathrm{#1}}
\newcommand{\mbb}[1]{\mathbb{#1}}

\newcommand{\gray}[1]{\textcolor{gray}{#1}}

\newcommand{\Sequence}[1]{\langle #1 \rangle}
\newcommand{\Set}[1]{\{ #1 \}}
\newcommand{\TreeSet}{\mcal{T}}

\newcommand{\InDeg}{\Delta^{-}}
\newcommand{\InEdges}{\delta^{-}}
\newcommand{\OutDeg}{\Delta^{+}}
\newcommand{\OutEdges}{\delta^{+}}

\newcommand{\SL}{\mcal{S}}
\newcommand{\ECnt}{\sigma}
\newcommand{\DES}{\mcal{D}}
\newcommand{\TG}{\mcal{G}}
\newcommand{\PH}{\mrm{PH}}
\newcommand{\PHS}{\mrm{PHS}}

\SetArgSty{textup}
\SetKwInOut{Input}{input}
\SetKwInOut{Output}{output}

\SetCommentSty{mycommfont}

\input{figs/fig_macro.tex}

\newcommand{\onappendix}[1]{} 

\usepackage[top=2cm, bottom=2cm, left=2cm, right=2cm, includefoot]{geometry}
\usepackage{authblk}
\usepackage{url}

\theoremstyle{definition}
\newtheorem{Problem}{Problem}
\theoremstyle{plain}
\newtheorem{definition}{Definition}

\newtheorem{theorem}{Theorem}
\newtheorem{lemma}{Lemma}
\newtheorem{corollary}{Corollary}
\theoremstyle{remark}

\makeatletter
\renewenvironment{proof}[1][\proofname]{\par
	\normalfont \topsep6\p@\@plus6\p@\relax
	\trivlist
	\item\relax
	{\itshape
		#1\@addpunct{.}}\hspace\labelsep\ignorespaces
}{%
	\popQED\endtrivlist\@endpefalse
}
\makeatother

\begin{document}

\title{Inferring Strings from Position Heaps in Linear Time}

\author{Koshiro~Kumagai}
\author{Diptarama~Hendrian}
\author{Ryo~Yoshinaka}
\author{Ayumi~Shinohara}

\date{}

\affil{Tohoku University, Japan}

\maketitle

\begin{abstract}
Position heaps are index structures of text strings used for the string matching problem.
They are rooted trees whose edges and nodes are labeled and numbered, respectively.
This paper is concerned with variants of the inverse problem of position heap construction and gives linear-time algorithms for those problems.
The basic problem is to restore a text string from a rooted tree with labeled edges and numbered nodes.
In the variant problems, the input trees may miss edge labels or node numbers which we must restore as well.
\date{}
\end{abstract}

\section{Introduction} \label{sec:intro}

The string matching problem searches for occurrences of a pattern $P$ in a text $T$.
It has been widely studied for many years and many efficient algorithms have been proposed.
Those techniques can be classified into mainly two approaches.
The first one is to construct data structures from $P$ by preprocessing $P$.
For example, the Knuth-Morris-Pratt algorithm~\cite{kmp} constructs border arrays,
the Boyer-Moore method~\cite{bm} constructs suffix tables,
and the Z-algorithm~\cite{gusfield1997algorithms} constructs prefix tables which is the dual notion of suffix tables.
The other approach is preprocessing $T$ to create indexing structures, such as suffix trees~\cite{stree}, suffix arrays~\cite{Manber1993}, LCP arrays~\cite{Manber1993}, suffix graphs~\cite{dawg}, compact suffix graphs~\cite{dawg2}, and position heaps~\cite{position-heap}.
Indexing structures are advantageous when searching for many different patterns in a text.

The reverse engineering of those data structures has also been widely studied.
Studying reverse engineering deepens our insight into those data structures.
For example, it may enable us to design an algorithm generating indexing structures with specific structural characteristics, which should be useful for verifying other software processing them.
The early studies targeted border arrays~\cite{Duval2005,duval2009efficient,Franek2002}.
Later, Cl{\'{e}}ment et al.~\cite{Clement2009} proposed a linear time algorithm for inferring strings from prefix tables.
Those data structures are produced by preprocessing patterns.
The reverse engineering for indexing structures has been studied for
suffix arrays~\cite{Bannai2003,duval2002words}, LCP arrays~\cite{Karkkainen2017}, suffix graphs~\cite{Bannai2003}, and suffix trees~\cite{CAZAUX20149,prior-bi-suffix-tree,prior-suffix-tree}.
The techniques used in~\cite{prior-bi-suffix-tree} and~\cite{prior-suffix-tree} involve finding an Eulerian cycles on a graph modifying an input tree.

In this paper, we discuss the reverse engineering of another type of indexing structures, called \emph{position heaps}~\cite{position-heap,online-position-heap}.
The {position heap} of a string $T$ is a rooted tree with labeled edges and numbered nodes.
Actually, Ehrenfeucht et al.~\cite{position-heap} and Kucherov~\cite{online-position-heap} gave different definitions of position heaps.
By either definition, position heaps can be constructed in linear time online assuming the alphabet size to be constant.
In addition, we can find all occurrence positions of a pattern $P$ in $O(|P|^2 + k)$ time, where $k$ is the output size.
Moreover, by augmenting position heaps with additional data structures, we can improve the searching time to $O(|P| + k)$.

We consider the following four types of reverse engineering of Kucherov's position heaps~\cite{online-position-heap}.
The first problem is to restore a source text $T$ from an input edge-labeled and node-numbered rooted tree so that the input should be the position heap $\PH(T)$ of $T$.
While this problem allows at most one solution, the other problems may have many possible solutions.
In the second problem, input trees miss edge labels.
In the third problem, input trees miss node numberings.
Instance trees of the fourth problem miss both edge labels and node numberings but have potential \emph{suffix links} among nodes, which play an important role in the construction of position heaps.
We show that all the problems above can be solved in linear time in the input size.
Among those, we devote the most pages to the third problem.
We reduce the problem to finding a special type of Eulerian cycle over the input tree augmented with suffix links.
By showing the problem of finding an Eulerian cycle of this special type is linear-time solvable, we conclude that restoring a text from a position heap without node numbers is linear-time solvable.
This can be seen analogous to the techniques used in~\cite{prior-bi-suffix-tree} and~\cite{prior-suffix-tree} for the suffix tree reverse engineering.
In addition, we present formulas for counting the number of possible text strings, which can be computed in polynomial time.
Moreover, we show efficient algorithms for enumerating all possible text strings in output linear time.
\section{Preliminaries} \label{sec:preliminaries}

Let $\Sigma$ be a finite alphabet and let the size of $\Sigma$ be constant.
For a string $w$ over $\Sigma$, the length of $w$ is denoted by $|w|$.
The \emph{empty string} $\varepsilon$ is the string of length 0.
Throughout this paper, strings are 1-indexed.
For $1 \leq i \leq j \leq |w|$, we let $w[i]$ be the $i$-th letter of $w$, and $w[i:j]$ be the substring of $w$ which starts at position $i$ and ends at position $j$.
In particular, we denote $w[i:|w|]$ by $w[i:]$ and $w[1:j]$ by $w[:j]$.
The concatenation of two strings $s$ and $t$ is denoted by $st$.

Let $\mathbb{N}_0$ and $\mathbb{N}_1$ be the set of natural numbers including and excluding 0, respectively.
We denote the cardinality of a set $X$ by $|X|$.

\subsection{Graphs}

A \emph{directed multigraph} $G$ is a tuple $(V, E, \Gamma)$ where $V$ is the node set, $E \subseteq V \times V$ is the edge set, and $\Gamma \colon E \rightarrow \mathbb{N}_1$ gives each edge its multiplicity.
The \emph{head} and the \emph{tail} of an edge $(u,v) \in E$ are $v$ and $u$, respectively.
This paper disallows self-loops: $(v,v) \notin E$ for any $v \in V$.
When $\Gamma(e) = 1$ for all $e \in E$, $G$ is called a \emph{directed graph} and is simply denoted by $(V, E)$.
An \emph{edge-labeled multigraph} is a tuple $(V, E, \Gamma, \Psi)$ where $\Psi \colon E \to \Sigma$ for an alphabet $\Sigma$.
A sequence $p = \Sequence{ e_1, \dots, e_\ell }$ of edges is called a \emph{$v_0$--$v_\ell$ path} if there are $v_0, \dots, v_\ell \in V$ such that $e_i = (v_{i-1}, v_i)$ for all $i \in \Set{1, \ldots, \ell}$.
Note that, the same node may occur more than once in a path in this paper.
We call $p$ a \emph{$v_0$-cycle} when $v_0 = v_\ell$.
For a $t$--$u$ path $p_1$ and a $u$--$v$ path $p_2$, we denote by $p_1 \cdot p_2$ the concatenation of $p_1$ and $p_2$, which will be a $t$--$v$ path.
By extending the domain of $\Psi$ to sequences of edges, we define
the \emph{path label} $\Psi(p)$ of $p$ to be the string $\Psi(e_1) \cdots \Psi(e_\ell)$.
When there exists just one $v_0$--$v_\ell$ path, we call its label the $v_0$--$v_\ell$ path label and denote it by $\Psi((v_0,v_\ell)) \in \Sigma^*$.

A directed graph $G$ is a \emph{$t$-rooted tree} ($t \in V$) if there exists exactly one $t$--$v$ path for all $v \in V$.
We call $t$ the \emph{root} of $G$.
Similarly, $G$ is a \emph{$t$-oriented tree} if there exists exactly one $v$--$t$ path for all $v \in V$.
We call $t$ the \emph{sink} of $G$.
For a $t$-rooted tree $G = (V, E)$, if $(u, v) \in E$, then $u$ is the \emph{parent} of $v$ and $v$ is a \emph{child} of $u$.
For two nodes $u, v \in V$ such that a $u$--$v$ path exists, $v$ is a \emph{descendant} of $u$, and $u$ is an \emph{ancestor} of $v$.
The \emph{depth} of $v$ is the length of the unique path from the root to $v$.
We denote the set of all descendants of $v$ as $\DES_G(v)$.

Two directed multigraphs $G = (V, E, \Gamma)$ and $G' = (V', E', \Gamma')$ are \emph{isomorphic}, denoted by $G \equiv G'$, if there is a bijection $\phi$ over $V$ such that $V' = \phi(V)$, $E' = \{\,(\phi(u),\phi(v)) \mid (u,v) \in E\,\}$, and $\Gamma'((\phi(u),\phi(v))) = \Gamma((u,v))$.
The definition of isomorphism is naturally extended and applied for edge-labeled directed multigraphs.
When $G$ is a rooted tree, we can verify $G \equiv G'$ in linear time.
If $V' = V$ and $G'$ is a $t$-oriented tree, then $G'$ is a \emph{$t$-oriented spanning tree} of $G$.

Let $G = (V, E, \Gamma)$ be a directed multigraph.
For a node $v \in V$, $\InEdges_G(v)$ and $\OutEdges_G(v)$ are the sets of edges whose heads and tails are $v$, respectively.
We denote the sum of the multiplicities of edges contained in $\InEdges_G(v)$ and $\OutEdges_G(v)$ by $\InDeg_G(v) = \sum_{e \in \InEdges_G(v)} \Gamma(e)$ and $\OutDeg_G(v) = \sum_{e \in \OutEdges_G(v)} \Gamma(e)$, respectively.
A cycle $p$ is \emph{Eulerian} when $p$ contains $e$ just $\Gamma(e)$ times for all $e \in E$.
We also call a directed multigraph \emph{Eulerian} if it has an Eulerian cycle.
It is well-known that $G$ is Eulerian if and only if $G$ is connected and $\InDeg_G(v) = \OutDeg_G(v)$ for all $v \in V$~\cite{eulerian-proof}.
Therefore, we can check whether $G$ is Eulerian in $O(|V| + |E|)$ time.
We often drop the subscript $G$ from $\DES_G$, $\OutEdges_G$, $\InDeg_G$ etc.\ when $G$ is clear from the context.

\subsection{Position heaps}

A position heap is an index structure with which one can efficiently solve the pattern matching problem.
In this paper, we follow Kucherov's definition~\cite{online-position-heap}.
Let $T$ be a string of length $n$ ending with a unique letter, i.e., $T[i] \neq T[n]$ for all $i \in \Set{1, \ldots, n-1}$.
The position heap $\PH(T)$ of $T$ is an edge-labeled rooted tree $(V, E, \Psi)$ defined as follows.
Let $h_0$ be $\varepsilon$, and $h_i$ be the shortest prefix of $T[i:]$ not contained in $\Set{h_0, \ldots, h_{i-1}}$ for all $i \in \Set{1, \ldots, n}$.
Since $T$ ends with a unique letter, $T[i:] \ne h_j$ for any $j < i$, and thus $h_i$ is always defined.
Then, define $V = \Set{0, \ldots, n}$, $E = \{(i, j) \mid h_i c = h_j \text{ for some } c \in \Sigma\}$, and $\Psi((i, j)) = c$ if $h_i c = h_j$.
Clearly, a position heap is $0$-rooted and $h_i$ is the $0$--$i$ path label for all $i \in \Set{0, \ldots, n}$.
Moreover, we have $i \le j$ if node $i$ is an ancestor of node $j$.
We call $T$ the \emph{source text} of $\PH(T)$.
Kucherov showed that one can determine whether a pattern $P$ occurs in $T$ in $O(|P|^2)$ time using $\PH(T)$.
Moreover, we can determine it in $O(|P|)$ time with auxiliary data structures.

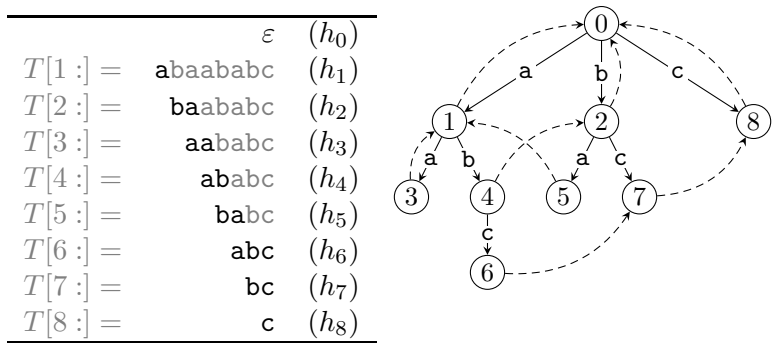
\begin{figure}[t]
  \centering
  \begin{tikzpicture}
    \MakeTree{{a/1/{a/3/,b/4/{c/6/}},b/2/{a/5/,c/7/},c/8/}}
    \foreach \x/\y/\z in {%
      -1/left/,-1-1/left/-1,-1-2/left/-2,-1-2-1/right/-2-2,%
      -2/right/,-2-1/right/-1,-2-2/right/-3,-3/right/%
    } {\path [suffix link] (root\x) to [bend \y] (root\z);}
    \node [anchor=north east] at ([xshift=-1mm] root.north -| root-1-1.west) {
      \begin{tabular}{rrc} \hline
                          & $\varepsilon$            & $(h_0)$ \\
        \gray{$T[1:] = $} & \texttt{a\gray{baababc}} & $(h_1)$ \\
        \gray{$T[2:] = $} & \texttt{ba\gray{ababc}}  & $(h_2)$ \\
        \gray{$T[3:] = $} & \texttt{aa\gray{babc}}   & $(h_3)$ \\
        \gray{$T[4:] = $} & \texttt{ab\gray{abc}}    & $(h_4)$ \\
        \gray{$T[5:] = $} & \texttt{ba\gray{bc}}     & $(h_5)$ \\
        \gray{$T[6:] = $} & \texttt{abc}             & $(h_6)$ \\
        \gray{$T[7:] = $} & \texttt{bc}              & $(h_7)$ \\
        \gray{$T[8:] = $} & \texttt{c}               & $(h_8)$ \\ \hline
      \end{tabular}
    };
  \end{tikzpicture}
  \caption{$\PHS(\texttt{abaababc})$ (dashed arrows are suffix links)}
  \label{fig:position-heap}
\end{figure}

In Kucherov's algorithm for constructing position heaps, the mapping $\SL \colon V \setminus \{0\} \rightarrow V$ called \emph{suffix links} plays an important role.
It is defined by $\SL(i) = j$ such that $h_i = c h_j$ for some $c \in \Sigma$ for $i > 0$.
The suffix links are well-defined.
It is clear that the depth of node $i$ is the depth of node $\SL(i)$ plus 1.
We often treat $\SL$ as a subset of $V \times V$.
We denote the position heap augmented with its suffix links by $\PHS(T) = (V, E, \Psi, \SL)$.
Figure~\ref{fig:position-heap} shows $\PHS(T)$ for $T = \texttt{abaababc}$.

\subsection{Problem definitions}

In this paper, we consider the following inverse problems of position heap construction.
The first problem is inferring the source text $T$ from a position heap.
\begin{Problem}[Inferring source texts from node-numbered edge-labeled trees] \label{prob:first}\ \\
  \tbf{Input}: An edge-labeled rooted tree $(V, E, \Psi)$ with $V=\{0,\dots,|V|-1\}$.\\
  \tbf{Output}: A string $T$ such that $\PH(T) = (V, E, \Psi)$ if such $T$ exists. Otherwise, ``invalid''.
\end{Problem}
We will also consider the problem where edge labels are missing.
\begin{Problem}[Inferring source texts from node-numbered trees] \label{prob:second}\ \\
  \tbf{Input}: A rooted tree $(V, E)$ with $V=\{0,\dots,|V|-1\}$.\\
  \tbf{Output}: A string $T$ such that $\PH(T) = (V, E, \Psi)$ for some $\Psi$ if such $T$ exists. Otherwise, ``invalid''.
\end{Problem}
The third problem is inferring source texts $T$ from trees whose nodes are not numbered but edges are labeled.
\begin{Problem}[Inferring source texts from edge-labeled trees] \label{prob:third}\ \\
  \tbf{Input}: An edge-labeled rooted tree $(V, E, \Psi)$.\\
  \tbf{Output}: A string $T$ such that $\PH(T) \equiv (V, E, \Psi)$ if such $T$ exists. Otherwise, ``invalid''.
\end{Problem}
In the end, we will address the problem where the input trees miss both node numbers and edge labels but have potential suffix links.
\begin{Problem}[Inferring source texts from trees with links] \label{prob:fourth}\ \\
  \tbf{Input}: A pair $(G, \SL)$ of a rooted tree $G = (V, E)$ and a partial map $\SL \colon V \rightarrowtail V$.\\
  \tbf{Output}: A string $T$ such that $\PHS(T) \equiv (V, E, \Psi, \SL)$ for some $\Psi$ if such $T$ exists. Otherwise, ``invalid''.
\end{Problem}

Figure~\ref{fig:given-graph} shows examples of instances of Problem~\ref{prob:second} and \ref{prob:third}
and Figure~\ref{fig:third-answer} shows all possible answers for the instance of Figure~\ref{fig:given-graph}(b).
\section{Proposed algorithms} \label{sec:proposal}

\subsection{Inferring source texts from node-numbered edge-labeled trees}

Solving Problem~\ref{prob:first} is easy.
Given an edge-labeled tree $(V, E, \Psi)$ where $V = \Set{0, \ldots, n}$,
let $h_i$ be the $0$--$i$ path label on $G$ for every $i \in V$.
If the input is the position heap of some string $T$, it must hold $T[i] = h_i[1]$.
Therefore, by DFS on $G$ remembering the initial letter of each path label,
 we can construct the candidate string $T$ in linear time.
Then, we can verify whether $\PH(T) = (V, E, \Psi)$ in linear time, since the position heap of $T$ can be constructed in linear time~\cite{online-position-heap}.
\begin{theorem} \label{th:first-problem}
  Problem~\ref{prob:first} is solvable in linear time.
\end{theorem}

\subsection{Inferring source texts from node-numbered trees}

\begin{figure}[t]
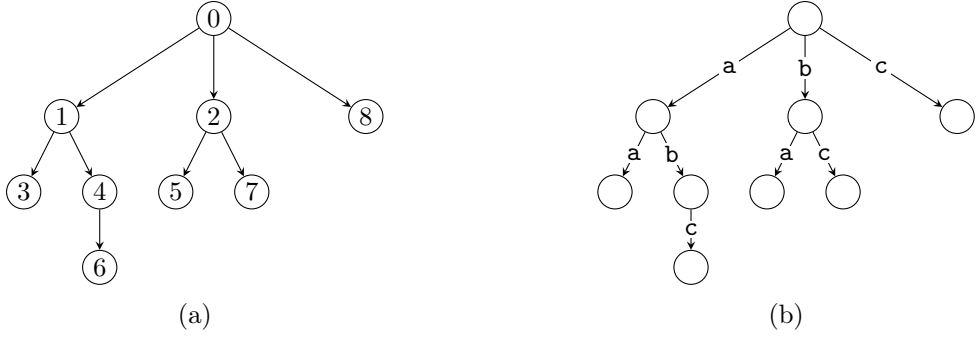

  \centering
  \begin{minipage}{.45\textwidth}
    \centering
    \tikz \MakeTree{{/1/{/3/,/4/{/6/}},/2/{/5/,/7/},/8/}};
    \subcaption{}
    \label{fig:given-graph:second}
  \end{minipage}
  \begin{minipage}{.45\textwidth}
    \centering
    \tikz \MakeTree[]{{a//{a//,b//{c//}},b//{a//,c//},c//}};
    \subcaption{}
    \label{fig:given-graph:third}
  \end{minipage}
  \caption{Examples of inputs to instances of (a) Problem~\ref{prob:second} and (b) Problem~\ref{prob:third}.}
  \label{fig:given-graph}
\end{figure}

\begin{figure}[t]
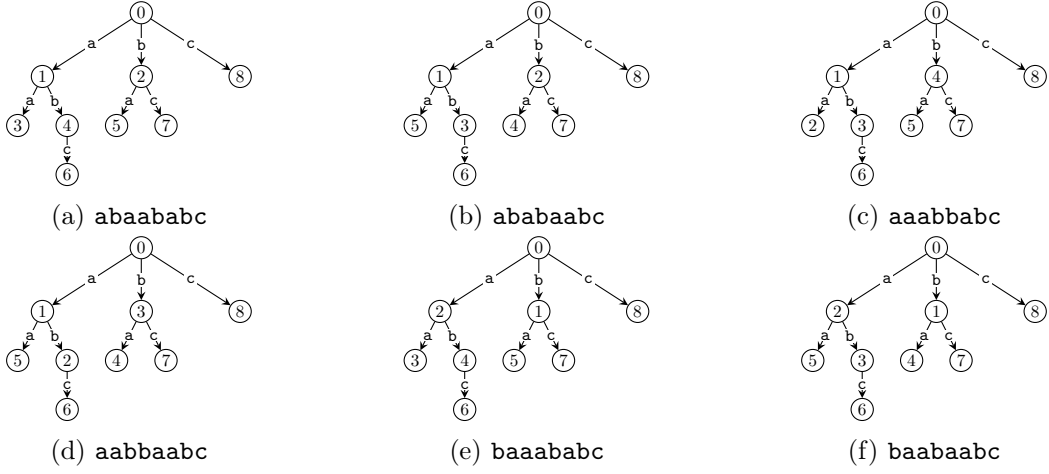

  \centering
  \begin{minipage}{.3\textwidth}\centering
    \tikz[scale=.65, transform shape] \MakeTree{{a/1/{a/3/,b/4/{c/6/}},b/2/{a/5/,c/7/},c/8/}};
    \subcaption{\texttt{abaababc}}
  \end{minipage}
  \begin{minipage}{.3\textwidth}\centering
    \tikz[scale=.65, transform shape] \MakeTree{{a/1/{a/5/,b/3/{c/6/}},b/2/{a/4/,c/7/},c/8/}};
    \subcaption{\texttt{ababaabc}}
  \end{minipage}
  \begin{minipage}{.3\textwidth}\centering
    \tikz[scale=.65, transform shape] \MakeTree{{a/1/{a/2/,b/3/{c/6/}},b/4/{a/5/,c/7/},c/8/}};
    \subcaption{\texttt{aaabbabc}}
  \end{minipage}\\
  \begin{minipage}{.3\textwidth}\centering
    \tikz[scale=.65, transform shape] \MakeTree{{a/1/{a/5/,b/2/{c/6/}},b/3/{a/4/,c/7/},c/8/}};
    \subcaption{\texttt{aabbaabc}}
  \end{minipage}
  \begin{minipage}{.3\textwidth}\centering
    \tikz[scale=.65, transform shape] \MakeTree{{a/2/{a/3/,b/4/{c/6/}},b/1/{a/5/,c/7/},c/8/}};
    \subcaption{\texttt{baaababc}}
  \end{minipage}
  \begin{minipage}{.3\textwidth}\centering
    \tikz[scale=.65, transform shape] \MakeTree{{a/2/{a/5/,b/3/{c/6/}},b/1/{a/4/,c/7/},c/8/}};
    \subcaption{\texttt{baabaabc}}
  \end{minipage}
  \caption{All possible answers to Problem~\ref{prob:third} when the graph in Figure~\ref{fig:given-graph:third} is given.}
  \label{fig:third-answer}
\end{figure}

Figure~\ref{fig:given-graph:second} shows an input to an instance of Problem~\ref{prob:second}.
The following procedure solves Problem~\ref{prob:second}.
We label the outgoing edges of the root with arbitrary but distinct letters of $\Sigma$.
Then, we construct an output candidate $T$ following the method for Problem~\ref{prob:first} in the previous subsection.

\begin{theorem} \label{th:second-problem}
  Problem~\ref{prob:second} is solvable in linear time.
\end{theorem}

There can be many correct outputs for input unless it is invalid.
The number of possible source texts to output equals the number of how to attach the labels to edges from the root $r$.
Since the number of letters that appear in $T$ equals $\OutDeg(r)$, the number of possible texts is $|\Sigma|! \: / \: \bigl(|\Sigma| - \OutDeg(r)\bigr)!$.
One can enumerate such $T$ in output linear time because one can enumerate all $\OutDeg(r)$-permutations of $\Sigma$ in output linear time~\cite{sedgewick1977permutation}.

\subsection{Inferring source texts from edge-labeled trees}

Compared to the previous two problems, solving Problem~\ref{prob:third} in linear time requires more elaborate arguments.
In this subsection, we assume that two distinct outgoing edges of a node have different labels, since otherwise obviously the input cannot be extended to a position heap.
We will investigate the structural properties of position heaps augmented with the suffix links,
 and see that the text $T$ will appear as the label of a path with a specific property over $\PHS(T)$.

\begin{lemma} \label{lem:node-position}
	Let $\PHS(T) = (V, E, \Psi, \SL)$ with $V=\{0,\dots,n\}$.
	We have $\SL(n)=0$ and $i + 1 \in \DES(\SL(v))$ for all $i \in V \setminus \{0, n\}$.
\end{lemma}

\begin{proof}
	We show the lemma by induction on the depth of node $i$.
	When the depth of node $i$ is $1$, $\SL(i)$ is the root $0$, which is an ancestor of every node including $i+1$.
	Note that the depth of node $n$ is $1$ since $T$ ends with a unique letter.
	Suppose the depth of node $i < n$ is two or more.
	In this case, let the $0$--$i$ path label $h_i$ be $a w b$ for some $a,b \in \Sigma$ and $w \in \Sigma^*$.
	Let $j$ be the parent of $i$, for which $h_j = aw$.
	Let $i_s = \SL(i)$ and $j_s = \SL(j)$, i.e., $h_{i_s} = wb$ and $h_{j_s} = w$.
	By the induction hypothesis, we have $j+1 \in \DES(j_s)$, i.e., $h_{j+1} = h_{j_s} w'$ for some $w' \in \Sigma^*$, which implies that $j+1 \ge j_s$.
	Together with the fact that $i > j$, we have $i+1 > j_s$.
	Since $h_i$ and $h_{i+1}$ are prefixes of $T[i:]$ and $T[i+1:]$, respectively,
	either $h_i[2:] = wb$ is a prefix of $h_{i+1}$ or the other way around.
	The fact $h_{j_s} = w$ and $i+1 > j_s$ implies that $wb$ is a prefix of $h_{i+1}$.
	That is, $h_{i_s}$ is a prefix of $h_{i+1}$, which means $i+1 \in \DES(i_s)$.
  \qed
\end{proof}
Hereafter, by a \emph{path/cycle} of $\PHS(T)= (V, E, \Psi, \SL)$, we mean a path/cycle of $(V,E \cup \SL)$.
We call elements of $E \cup \SL$ \emph{arcs} while reserving the term {edges} for elements of $E$.
From Lemma~\ref{lem:node-position}, for all $i \in \Set{1, \ldots, n-1}$, $\PHS(T)$ has a special $i$--$(i+1)$ path which starts with the suffix link followed by zero or some number of edges.
We define a cycle by concatenating all special $i$--$(i+1)$ paths.

\begin{definition}
For $\PHS(T) = (V, E, \Psi, \SL)$,
let $f_i = (i, \SL(i))$, $p_0$ the path from $0$ to $1$, and $p_i$ the path from $\SL(i)$ to $i+1$ for $i > 0$.
The \emph{$T$-trace cycle} of $\PHS(T)$ is the sequence $p_0 \cdot f_1 \cdot p_1 \cdots f_{n-1} \cdot p_{n-1} \cdot f_n$.
\end{definition}
Figure~\ref{fig:trace-cycle} shows the $T$-trace cycle of $\PHS(T)$ for $T=\texttt{abaababc}$.
Note that the $T$-trace cycle is a cycle in the graph $(V,E \cup \SL)$, where each element of $\SL$ appears exactly once.
Since following an edge from $E$ and a suffix link from $\SL$ increases and decreases the depth by one, respectively, the total numbers of occurrences of edges and suffix links in the $T$-trace cycle should be balanced.
That is, the $T$-trace cycle contains exactly $n$ occurrences of edges from $E$.
The following lemma explains why we call the cycle $T$-trace cycle.

\begin{figure}[t]
  \begin{minipage}{0.48\textwidth}
  \centering
  \begin{tikzpicture}
    \begin{scope}[edge from parent/.style={draw=none}, black!80]
      \MakeTree{{/1/{/3/,/4/{/6/}},/2/{/5/,/7/},/8/}}
    \end{scope}

    \draw [preaction={draw=white, line width=3pt}, thick]
      (root.center) to (root.-150)
                    to (root-1.30)
                    to [bend left=60] (root-1.60)
      (root.180) to [bend left] (root.-90)
                 to (root-2.90)
                 to [bend right=60] (root-2.60)
      (root.-60) to [bend right=60] (root.-135)
                 to (root-1.15)
                 to [bend right=15] (root-1.-120)
                 to (root-1-1.60)
                 to [bend left=60] (root-1-1.90)
      (root-1.-150) to [bend left=60] (root-1.-60)
                    to (root-1-2.120)
                    to [bend right=60] (root-1-2.60)
      (root-2.180) to [bend left=60] (root-2.-120)
                   to (root-2-1.60)
                   to [bend left=60] (root-2-1.120)
      (root-1.0) to [bend right=30] (root-1.-45)
                 to (root-1-2.90)
                 to (root-1-2.-90)
                 to (root-1-2-1.90)
                 to [bend right=60] (root-1-2-1.0)
      (root-2-2.-120) to [bend left=15] (root-2-2.0)
      (root-3.-120) to [bend right=15] (root-3.120)
      ;

    \draw [preaction={draw=white, line width=3pt}, thick, densely dashed]
      (root-1.60) to [bend left] (root.180)
      (root-2.60) to [bend right] (root.-60)
      (root-1-1.90) to [bend left] (root-1.-150)
      (root-1-2.60) to [bend left] (root-2.180)
      (root-2-1.120) to [bend right] (root-1.0)
      (root-1-2-1.0) to [bend right] (root-2-2.-120)
      (root-2-2.0) to [bend right] (root-3.-120)
      ;

    \draw [preaction={draw=white, line width=3pt}, thick, densely dashed, -latex] (root-3.120) to [bend right] (root.0);
    \draw [fill=black] (root.center) circle (2pt);
  \end{tikzpicture}
  \caption{The $T$-trace cycle of $\PHS(T)$ with $T=\texttt{abaababc}$, which is an answer to the input graph in Figure~\ref{fig:given-graph:third}. Dashed lines represent suffix links.}
  \label{fig:trace-cycle}
  \end{minipage}
  \hfill
  \begin{minipage}{0.5\textwidth}
  \centering
  \begin{tikzpicture}
    \begin{scope}[efp/.style={draw=none}]
      \MakeTree[]{{//{//,//{//}},//{//,//},//}}
    \end{scope}

    \foreach \x/\y in {%
      /-2,-1/-1-1,-1/-1-2,-1-2/-1-2-1,-2/-2-1%
    } {\path [efp] (root\x) to (root\y);}

    \foreach \x/\y in {%
      /-1,-1/-1-2%
    } {\path [efp, double] (root\x) to (root\y);}

    \foreach \x/\y/\z in {%
      -1/left/,-1-1/left/-1,-1-2/left/-2,-1-2-1/right/-2-2,%
      -2/right/,-2-1/right/-1,-2-2/right/-3,-3/right/%
    } {\path [suffix link] (root\x) edge [bend \y] (root\z);}
  \end{tikzpicture}
  \caption{
  The trace graph of the input graph in Figure~\ref{fig:given-graph:third}. The multiplicities of doubled edges are 2 and the others are 1. Dashed arrows show suffix links.}
  \label{fig:trace-graph}
  \end{minipage}
\end{figure}

\begin{lemma} \label{lem:trace-text}
	Let $e \in E$ be the $i$-th occurrence of an edge in the $T$-trace cycle of $\PHS(T)$.
  Then $\Psi(e) = T[i]$.
\end{lemma}

\begin{proof}
	Suppose the $i$-th edge $e = (u,v)$ in the $T$-trace cycle $p = p_0 \cdot f_1 \cdots p_{n-1} \cdot f_n$ occurs in the $p_j$ segment.
	In other words, $p$ can be written as $p' \cdot (u,v) \cdot p''$, where $p'$ contains $j$ suffix links and $i-1$ edges.
	Then, the depth of $v$ is $i-j$.
	Moreover, the edge $e$ is on the path from the root to the node $j+1$, whose label is a prefix of $T[j+1:]$.
  That is, $\Psi(e)$ is the $(i-j)$-th letter of $T[j+1:]$.
	Hence, $\Psi(e) = T[(j+1)+(i-j)-1] = T[i]$.
  \qed
\end{proof}

Lemma~\ref{lem:trace-text} allows us to spell $T$ by following the $T$-trace cycle without referring to node numbers.
To solve Problem~\ref{prob:third}, we will construct the $T$-trace cycle of $\PHS(T) \equiv (V, E, \Psi, \SL_G)$ for some $T$ from the input graph $G = (V, E, \Psi)$.
For this end, we first reconstruct the suffix links $\SL$.

\begin{lemma} \label{lem:suffix-link}
	From an edge-labeled rooted tree $G = (V, E, \Psi)$, one can uniquely construct $\SL$ in linear time
	 such that $\PHS(T) \equiv (V, E, \Psi, \SL)$ for some $T$ if any exist.
\end{lemma}

\begin{proof}
	We recover the suffix links of nodes from shallower to deeper.
	Let $r$ be the root of $G$.
	From the definition of suffix links, we have $\SL(v) = r$ for every node of depth $1$.
	For $e = (u, v) \in E$ with $\Psi(e) = c$, we assume $\SL(u)$ has already been determined.
	Let $a w$ be the $r$--$u$ path label where $a \in \Sigma$ and $w \in \Sigma^*$.
	The $r$--$\SL(u)$ path label is $w$ and the $r$--$v$ path label is $a w c$.
	Therefore, the $r$--$\SL(v)$ path label is $w c$.
	Hence, an edge $(\SL(u), \SL(v))$ labeled $c$ exists.
	So, for the node $t \in V$ such that $(\SL(u),t) \in E$ and $\Psi((\SL(u),t))=c$, we determine $\SL(v) = t$.
  \qed
\end{proof}

If we fail to give a suffix link to any of the nodes by the procedure described in the proof of Lemma~\ref{lem:suffix-link}, the answer to Problem~\ref{prob:third} is ``invalid''.

While the $T$-trace cycle contains just one occurrence of each suffix link, the numbers of occurrences of respective edges vary.
Actually, one can uniquely determine the multiplicity of each edge in the $T$-trace cycle from $G$.
\begin{lemma} \label{lem:edge-cnt}
	Let $\ECnt(e)$ be the number of occurrences of $e$ in the $T$-trace cycle for all $e \in E$.
  Then, it holds that
  \begin{equation}
    \ECnt(e) = 1 - \bigl|\Set{u \in V \mid \SL_G(u) = v}\bigr| + \sum_{e' \in \OutEdges_G(v)} \ECnt(e') \label{eq:edge-cnt}
  \end{equation}
  where $v$ is the head of $e$.
\end{lemma}
	Note that $\OutEdges_G(v)$ contains no suffix links of $\PHS(T)$.
\begin{proof}
	The $T$-trace cycle must include the same number of occurrences of arcs coming into and going out from node $v$.
	Since each suffix link occurs just once in the $T$-trace cycle, we obtain the lemma.
  \qed
\end{proof}
\begin{lemma}\label{lem:unique_multiplicity}
  The system of equations \eqref{eq:edge-cnt} in $\ECnt$ has a unique solution.
  Moreover, it can be computed in linear time.
\end{lemma}
\begin{proof}
One can uniquely determine the value of $\ECnt(e)$ inductively on the height of $e \in E$.
Then, the linear-time computation is obvious.
  \qed
\end{proof}

Let us call a cycle $p$ of $(V, E, \Psi, \SL)$ a \emph{legitimate cycle} if it is the $T$-trace cycle for some $T$.
Based on Lemmas~\ref{lem:edge-cnt} and~\ref{lem:unique_multiplicity}, we define the directed multigraph for which every legitimate cycle is Eulerian.
\begin{definition}[Trace graph]
	The \emph{trace graph} $\TG(G)$ of an edge-labeled tree $G = (V, E, \Psi)$ is a tuple $(V, E', \SL, \Gamma)$ where $E' = \Set{e \in E \mid \ECnt(e) > 0}$ and $\Gamma \colon E' \cup \SL \to \mbb{N}_1$ is defined by
	\begin{equation}
		\Gamma(e) = \begin{cases}
			1 & \text{if } e \in \SL, \\
			\ECnt(e) & \text{if } e \in E',
		\end{cases} \nonumber
	\end{equation}
	where $\SL$ and $\ECnt$ are given in Lemmas~\ref{lem:suffix-link} and~\ref{lem:unique_multiplicity}, respectively.
\end{definition}

Figure~\ref{fig:trace-graph} shows the trace graph of Figure~\ref{fig:given-graph:third}.
The doubled arrows have multiplicity 2 and the others have 1.
The dashed arrows are suffix links.

From the definition, it is obvious that the $T$-trace cycle is an $r$-Eulerian cycle of $\TG(G)$ where $r$ is the root of $G$.
However, not every Eulerian cycle of $\TG(G)$ can be a legitimate cycle.
Recall that in the definition of the $T$-trace cycle, the suffix link of every node $u$ proceeds all outgoing edges of $u$.
We say that an Eulerian cycle $p$ of $\TG(G)$ \emph{respects} $\SL$
 if no edges of $ \OutEdges_G(u)$ occur before $(u,\SL(u))$ in $p$.

\begin{lemma} \label{lem:reduction}
  A cycle $p$ is an $r$-Eulerian cycle respecting $\SL$ if and only if $p$ is the $T$-trace cycle of some $T$.
\end{lemma}
\begin{proof}
  ($\Longleftarrow$)
  By definition.

  ($\Longrightarrow$)
  Let $n = |V|$ and $r$ be the root of $G$.
  Let $p_i$ and $f_{i+1}$ be the sequences of edges and the suffix links for $i=0,\dots,n-1$ so that $p=p_0 \cdot f_1 \cdot p_1 \dots p_{n-1} \cdot f_n$.
  Since $p$ ends at $r$ and only suffix links point to $r$, $p$ always ends with a suffix link.
  We define the bijection $\Lambda \colon V \rightarrow \Set{0, \ldots, n}$ such that $\Lambda(r) = 0$ and $\Lambda(s) = i$ if $f_i = (s, \SL(s))$ for all $s \in V \setminus \{r\}$.
  Let $s_i$ be the node such that $\Lambda(s_i) = i$.

  We first show $\Lambda(u) < \Lambda(v)$ for all $(u, v) \in E$ by induction on $\Lambda(v)$.
  Suppose the claim holds true for all $v$ such that $\Lambda(v) < i$.
  Then, we will show the claim holds for the edge whose head is $s_i$.
  If $|p_i| \ge 1$, the edge $(s_k,s_i)$ occurs just before $f_i=(s_i,\SL(s_i))$ in $p$.
  Since $p$ respects $\SL$, $f_k=(s_k,\SL(s_k))$ occurs before $(s_k,s_i)$.
  Thus, we have $k < i$.
  If $|p_i| = 0$, $f_{i-1}=(s_{i-1},s_i)$.
  Let the parents of $s_{i-1}$ and $s_i$ be $s_j$ and $s_k$, respectively.
  By the induction hypothesis, $j < i-1$.
  By the definition of $\SL$, $f_j = (s_j,s_k) \in \SL$.
  Since $p$ respects $\SL$, $f_k = (s_k, \SL(s_k))$ appears either before $f_j$ or right after $f_j$.
  That is, $k \le j+1$ holds.
  Therefore, $k < i$.

  Now, we define a string $T$ by $T[i] = \Psi(e_i)$ where $e_i$ is the $i$-th edge in $p$ for $i=1,\dots,n$,
  and define $h_i$ inductively to be the shortest prefix of $T[i:]$ which is not in $\{h_0,\dots,h_{i-1}\}$ where $h_0=\varepsilon$.
  We will show by induction on $i$ that for all $j \le i$, the $s_0$--$s_j$ path label $\Psi((s_0,s_j))$ is $h_j = T[j:x_j]$ where $x_j = |p_0 \dots p_{j-1}|$.
  This implies $(V, E, \Psi) \equiv \PH(T)$ when $i = n$.
  Then the constructed $\SL$ is the correct suffix links of $\PH(T)$ by Lemma~\ref{lem:suffix-link} and thus $p$ is the $T$-trace cycle.

  Let $g_i = \Psi((s_0,s_i))$.
  The claim clearly holds for $i=0$ by $g_0=h_0=\varepsilon$.
  Suppose the claim holds true for $i$.
  That is, $g_i = h_i = T[i:x_i]$ where $x_i = |p_0 \dots p_{i-1}|$.
  Let $u = \SL(s_i)$.
  By the definition of $\SL$, we have $\Psi((s_0,u)) = g_i[2:] = T[i+1:x_i]$.
  By the definition of $T$, $\Psi((u,s_{i+1})) = p_{i} = T[x_i+1:x_i+|p_{i}|] = T[x_i+1:x_{i+1}]$, where $x_{i+1} = |p_0 \dots p_i|$.
  By concatenating these two paths, we obtain $g_{i+1} = \Psi((s_0,s_{i+1})) = T[i+1:x_{i+1}]$.
  Since the labels of all proper ancestors of $s_{i+1}$ are at most $i$, all prefixes of $g_{i+1}$ appears in $\{g_0,\dots,g_i\}=\{h_0,\dots,h_i\}$.
  That is, $g_{i+1}$ is the least prefix of $T[i+1:]$ not in $\{h_0,\dots,h_i\}$, i.e., $g_{i+1}=h_{i+1}$.
\qed
\end{proof}
Therefore, to find a source text $T$, it is enough to find an $r$-Eulerian cycle over $(V,E,\SL,\Gamma)$ that respects $\SL$ where $r$ is the root.
We show that this problem can be solved in linear time on general graphs.
\onappendix{We present pseudo-codes of our algorithms for solving Problems~\ref{prob:third} and~\ref{prob:ecp} in Appendix.}
\begin{Problem}[The ECP (Eulerian cycle with priority edges) problem] \label{prob:ecp}\ \\
  \textbf{Input}: A tuple $(G,F,r)$ of a directed multigraph $G=(V,E,\Gamma)$, an edge subset $F \subseteq E$, and a start node $r \in V$
  such that $|F \cap \OutEdges(v)| \le 1$ for all $v \in V$
  and $\Gamma(e)=1$ for all $e \in F$.\\
  \textbf{Output}: An $r$-Eulerian cycle that respects $F$ if any. Otherwise, ``invalid''.
\end{Problem}
We call edges of $F$ \emph{priority edges}.
Without loss of generality, we may assume a node has a priority outgoing edge only if it has another outgoing edge.
If a node has only one outgoing edge and it has priority, then one can remove it from $F$ and make it a non-priority edge.
This does not affect possible solutions.
In what follows, we show how to solve the ECP problem in linear time.

First, let us review a linear-time algorithm for constructing an $r$-Eulerian cycle.
The following procedure gives a justification for the so-called BEST theorem~\cite{BEST-theorem,BEST-theorem-graph},
which counts the number of Eulerian cycles in a directed multigraph.
\begin{enumerate}
  \item Construct an arbitrary $r$-oriented spanning tree $H$ of $G$, \label{itm:spanning_tree}
  \item Starting from $r$, choose an arbitrary unused edge to follow next, except that an edge in $H$ can be chosen only when it is the only remaining choice, until we follow all the edges of $G$.
\end{enumerate}
This process guarantees to find an Eulerian cycle without getting stuck.
We modify this procedure so that the output shall respect $F$.
\begin{enumerate}
  \item Construct an arbitrary $t$-oriented spanning tree $H$ of $(V,E \setminus F)$,
  \item Starting from $r$, choose an arbitrary unused edge to follow next, except that
  \begin{itemize}
    \item choose an unused priority edge if the current node has any,
    \item an edge in $H$ can be chosen only when it is the only remaining choice,
  \end{itemize}
    until we follow all the edges of $G$.
\end{enumerate}

\begin{theorem} \label{th:third-problem}
  We can compute an answer to the ECP problem in linear time.
\end{theorem}

One can count the number of $r$-ECPs by modifying the BEST theorem formula\onappendix{, which is presented in Appendix}.
Letting $G' = (V, E \setminus F, \Gamma')$ with the restriction $\Gamma'$ of $\Gamma$ to $E \setminus F$, the number of $r$-ECPs is given as
\begin{equation}
	\OutDeg_{G'}(r) \cdot \prod_{v \in V} \frac{(\OutDeg_{G'}(v) - 1)!}{\prod_{e \in \OutEdges_{G'}(v) } \Gamma'(e)!} \cdot \sum_{(V,E') \in \TreeSet_{G'}(r)} \prod_{e \in E'} \Gamma'(e) \label{eq:ECP_count}
\end{equation}
where $\TreeSet_{G'}(r)$ is the set of $r$-oriented spanning trees of $G'$.
One can compute \eqref{eq:ECP_count} in polynomial time by the matrix-tree theorem~\cite{matrix-tree}.

\begin{theorem} \label{th:third-problem-count}
  We can calculate the number of $r$-ECPs in polynomial time.
\end{theorem}

One can also enumerate $r$-ECPs.
We have already described a linear-time nondeterministic algorithm to find an $r$-ECP.
Gabow and Myers proposed an algorithm~\cite{enumerate-spanning-tree} to enumerate spanning trees in output linear time.
By searching all the possible choices of the procedure,
we enumerate all the $r$-ECPs.

\begin{theorem} \label{th:third-problem-enumeration}
  We can enumerate $r$-ECPs in linear time per solution.
\end{theorem}

\begin{corollary} \label{col:third-problem}
	Problem~\ref{prob:third} is solvable in linear time.
	Moreover, one can count and enumerate all possible answers in polynomial time and output linear time, respectively.
\end{corollary}
\begin{proof}
	The first claim follows from Theorem~\ref{th:third-problem}.
	By Theorems~\ref{th:third-problem-count} and~\ref{th:third-problem-enumeration},
	it suffices to show that two distinct legitimate cycles $p$ and $p'$ over a trace graph give different source texts.
	Suppose $e$ and $e'$ are the first mismatch of $p$ and $p'$.
	Since choosing a suffix link is obligatory, $e \ne e'$ implies $e,e' \in E$.
	Since distinct edges with the same tail have distinct labels, $\Psi(e) \neq \Psi(e')$, and thus those two cycles spell different source texts.
\qed
\end{proof}

\subsection{Inferring source texts from trees with links} \label{seq:fourth-problem}

Instance trees of Problem~\ref{prob:fourth} miss both node numbers and edge labels but have possible suffix links.
This problem can be solved by combining ideas for solving Problems~\ref{prob:second} and~\ref{prob:third}.
We first label the outgoing edges of the root node with arbitrary distinct letters.
Then, the other edge labels are uniquely determined by the definition of suffix links, as long as the input is valid.
Now, the algorithm for Problem~\ref{prob:third} can be applied.
Similarly one can solve the counting and enumerating variants of Problem~\ref{prob:fourth}.

\begin{theorem} \label{th:fourth-problem}
  We can solve Problem~\ref{prob:fourth} in linear time.
  Moreover, one can count the number of output strings in polynomial time, and enumerate all output strings in linear time per each.
\end{theorem}
\section{Conclusion} \label{sec:conclusion}

We studied four types of reverse engineering problems on Kucherov's position heaps~\cite{online-position-heap}
and showed that all problems can be solved in linear time.
One can think of an even more restrictive variant, where the input tree has no edge labels, no node numbers, and no suffix links.
In this setting, we need to find ``valid'' suffix links, which seems a challenging task.

One can also study the reverse engineering problems of position heaps based on the definition by Ehrenfeucht et al.~\cite{position-heap}.
We conjecture that those problems can be solved by quite similar techniques presented in this paper.

Another interesting direction of future work is to study the reverse engineering of augmented position heaps~\cite{position-heap}.

\bibliographystyle{plain}
\bibliography{ref}

\end{document}